\documentclass[aps,reprint,twocolumn,superscriptaddress,citeautoscript,showpacs,amsmath,amssymb,amsfonts,floatfix,prl]{revtex4-1}
\usepackage{gensymb}
\usepackage{graphicx}
\usepackage{dcolumn}
\usepackage{bm}
\usepackage{times}
\usepackage{color}
\begin{document}
\title{Itinerant ferromagnetism in the As 4$p$ conduction band of Ba$_{0.6}$K$_{0.4}$Mn$_{2}$As$_{2}$ identified by x-ray magnetic circular dichroism}

\author{B. G. Ueland}
\email{bgueland@ameslab.gov, bgueland@gmail.com}
\affiliation{Ames Laboratory, U.S. DOE, Iowa State University, Ames, Iowa 50011, USA}
\affiliation{Department of Physics and Astronomy, Iowa State University, Ames, Iowa 50011, USA}

\author{Abhishek Pandey}
\affiliation{Ames Laboratory, U.S. DOE, Iowa State University, Ames, Iowa 50011, USA}
\affiliation{Department of Physics and Astronomy, Iowa State University, Ames, Iowa 50011, USA}

\author{Y. Lee}
\affiliation{Ames Laboratory, U.S. DOE, Iowa State University, Ames, Iowa 50011, USA}
\affiliation{Department of Physics and Astronomy, Iowa State University, Ames, Iowa 50011, USA}

\author{A. Sapkota}
\affiliation{Ames Laboratory, U.S. DOE, Iowa State University, Ames, Iowa 50011, USA}
\affiliation{Department of Physics and Astronomy, Iowa State University, Ames, Iowa 50011, USA}

\author{Y. Choi}
\affiliation{Advanced Photon Source, Argonne National Laboratory, Argonne, Illinois 60439, USA}

\author{D. Haskel}
\affiliation{Advanced Photon Source, Argonne National Laboratory, Argonne, Illinois 60439, USA}

\author{\mbox{R. A. Rosenberg}}
\affiliation{Advanced Photon Source, Argonne National Laboratory, Argonne, Illinois 60439, USA}

\author{J. C. Lang}
\affiliation{Advanced Photon Source, Argonne National Laboratory, Argonne, Illinois 60439, USA}

\author{B. N. Harmon}
\affiliation{Ames Laboratory, U.S. DOE, Iowa State University, Ames, Iowa 50011, USA}
\affiliation{Department of Physics and Astronomy, Iowa State University, Ames, Iowa 50011, USA}

\author{D. C. Johnston}
\affiliation{Ames Laboratory, U.S. DOE, Iowa State University, Ames, Iowa 50011, USA}
\affiliation{Department of Physics and Astronomy, Iowa State University, Ames, Iowa 50011, USA}

\author{A. Kreyssig}
\affiliation{Ames Laboratory, U.S. DOE, Iowa State University, Ames, Iowa 50011, USA}
\affiliation{Department of Physics and Astronomy, Iowa State University, Ames, Iowa 50011, USA}

\author{A. I. Goldman}
\affiliation{Ames Laboratory, U.S. DOE, Iowa State University, Ames, Iowa 50011, USA}
\affiliation{Department of Physics and Astronomy, Iowa State University, Ames, Iowa 50011, USA}

\date{\today}

\begin{abstract}
X-ray magnetic circular dichroism (XMCD) measurements on single-crystal and powder samples of Ba$_{0.6}$K$_{0.4}$Mn$_{2}$As$_{2}$ show that the ferromagnetism below $T_{\textrm{C}}\approx$~100~K arises in the As $4p$ conduction band.  No XMCD signal is observed at the Mn x-ray absorption edges.  Below $T_{\textrm{C}}$, however, a clear XMCD signal is found at the As $K$ edge which increases with decreasing temperature.  The XMCD signal is absent in data taken with the beam directed parallel to the crystallographic \textbf{c} axis indicating that the orbital magnetic moment lies in the basal plane of the tetragonal lattice. These results show that the previously reported itinerant ferromagnetism is associated with the As $4p$ conduction band and that distinct local-moment antiferromagnetism and itinerant ferromagnetism with perpendicular easy axes coexist in this compound at low temperature.
\end{abstract}

\maketitle

The search for unconventional superconductivity has resulted in the discovery of materials possessing rich, tunable, physical properties and has been the driving force for much recent theoretical and experimental research \cite{Anderson_1987, Johnston_1997, Maple_2008,  Stewart_2011}.  In particular, the high-temperature cuprate and Fe-pnictide superconductors show remarkable coupling between their lattice, charge, and magnetic degrees of freedom, which may be tuned by small alterations to their chemical structures through, for example, electron- or hole-doping by chemical substitution. \cite{Lynn_2009, Paglione_2010, Canfield_2010}.  These materials have provided testbeds for Hamiltonians based on the Heisenberg model and their applicability to both local-moment and itinerant magnets \cite{Johnston_1997, Lee_2006, Johnston_2010, Dai_2012}.  In particular, the magnetic excitation spectra obtained through inelastic neutron scattering experiments on the parent compounds of the 122-type Fe-pnictide superconductors ($A$Fe$_{2}$As$_{2}$, $A=$~Ba, Sr, Ca) have been described in terms of the $J_{1}$-$J_{2}$ Heisenberg model \cite{Johnston_2010, Dai_2012}.

Systems related to the 122 Fe-pnictides manifest interesting and exotic material properties.  For example, (Ba$_{1-x}$K$_{x}$)(Zn$_{1-y}$Mn$_{y}$)$_{2}$As$_{2}$ is a ferromagnetic semiconductor which may be useful for developing multilayer functional devices \cite{Zhao_2013}.  The system studied here, (Ba$_{0.6}$K$_{0.4}$)Mn$_2$As$_2$, shows local-moment antiferromagnetic (AFM) order below a N\'{e}el temperature of  $T_{\textrm{N}}=480$~K that coexists with half-metallic itinerant ferromagnetic (FM) order below a Curie temperature of $T_{\textrm{C}}\approx$~100~K \cite{Pandey_2013}.

BaMn$_2$As$_2$ crystallizes in the same body-centered tetragonal ThCr$_{2}$Si$_{2}$ lattice (space group $I4/mmm$) as the parent phases of the superconducting 122 Fe-pnictides \cite{Brechtel_1978}. The G-type (checkerboard-type) AFM insulating ground state of the $S=\frac{5}{2}$  Mn spins has an ordered moment of $3.88(4)~\mu_{\textrm{B}}/\textrm{Mn}$ below $T_{\textrm{N}} = 625(1)$~K that is directed along the \textbf{c} axis. \cite{Singh_2009,An_2009,Singh_2009b,Johnston_2011}. It has been demonstrated that BaMn$_{2}$As$_{2}$ can be driven from an insulating to metallic ground state upon the substitution of as little as 1.6\% K for Ba \cite{Pandey_2012} or by applied pressure \cite{Satya_2011}.  Nevertheless, Ba$_{1-x}$K$_{x}$Mn$_{2}$As$_{2}$ retains the G-type Mn local-moment AFM order of its parent insulating compound \cite{Singh_2009b,Pandey_2012,Lamsal_2013}.  The AFM order is quite robust, persisting for substitutions up to at least $x=0.4$ with only nominal changes in $T_{\textrm{N}}$ and the ordered moment \cite{Lamsal_2013}.  The partial substitution of K for Ba modifies the electronic structure by hole-doping, and spin-polarized electronic structure calculations and angle-resolved photoemission spectroscopy measurements show that increasing the K doping results in a hole pocket that crosses the Fermi energy \cite{Pandey_2012}.

Perhaps most surprising is the observation of a novel ground state in Ba$_{1-x}$K$_{x}$Mn$_{2}$As$_{2}$ for $0.16\leq x\leq0.4$, where magnetization, nuclear magnetic resonance (NMR), and magnetic field dependent neutron diffraction measurements revealed the onset of itinerant FM below $T_{\textrm{C}} \approx$ 100~K coexisting with the Mn local-moment AFM order \cite{Pandey_2013, Bao_2012}.  Whereas the moment associated with the AFM order is directed along the unique \textbf{c} axis of the tetragonal structure, magnetization data indicate that the FM ordered moments ($\approx0.45(1)\mu_{\textrm{B}}/\textrm{formula unit}$ for x = 0.4)  are aligned perpendicular to this direction, in the \textbf{ab} plane \cite{Pandey_2013, Bao_2012}.

It has been proposed that the FM order in Ba$_{0.6}$K$_{0.4}$Mn$_{2}$As$_{2}$~arises from the complete spin polarization of the doped holes in the conduction band (half-metallic ferromagnetism) \cite{Pandey_2013}.  Nevertheless, it is reasonable to question whether the observed FM ordering in Ba$_{0.6}$K$_{0.4}$Mn$_{2}$As$_{2}$ arises from the canting of the ordered Mn moments. The above mentioned NMR and neutron diffraction measurements argue against this explanation \cite{Pandey_2013}, however, recent first-principles calculations suggest that canting of the Mn moments may be induced by a double-exchange interaction \cite{Glasbrenner_2013}.

Here, we use the element specificity of x-ray magnetic circular dichroism (XMCD) through measurements on single-crystal and powder samples of Ba$_{0.6}$K$_{0.4}$Mn$_{2}$As$_{2}$ to show that FM ordering occurs in the As $4p$ conduction band.  We find a strong peak in the XMCD signal at the As $K$ x-ray absorption edge, but a featureless XMCD signal  across the  Mn $L_{2}$ and $L_{3}$  absorption edges.  The noise in the signal observed at the Mn $L_{2}$ and $L_{3}$ absorption edges corresponds to a limit of detection of $0.02\mu_{\textrm{B}}/{\textrm{Mn}}$ and is an order of magnitude smaller than the FM moment determined from bulk magnetization measurements \cite{Pandey_2013}.  The XMCD signal at the As $K$ edge appears at $T \approx T_{\textrm{C}}$ due to FM ordering in the As $4p$ conduction band, and its intensity increases with decreasing temperature, in agreement with previous bulk magnetization measurements \cite{Pandey_2013}.  Our measurements on single-crystal samples further demonstrate that the FM moments lie in the \textbf{ab} plane.

The signal from XMCD measurements is obtained from the differential absorption of circularly-polarized x-rays when the helicity of the incident light is oriented parallel or antiparallel to the magnetization direction of a FM material. During the measurement, the x-ray energy is tuned through the absorption edges of the constituent elements \cite{Lovesey_1996, Lang_2003, Thole_1992, Carra_1993}, providing the elemental specificity.  The circularly-polarized photons induce a transition from occupied core states (e.g.\ $1s$ or $2p$ ground states for $K$ or $L$ edges, respectively) to unoccupied states with energies $E~\geq~E_{\textrm{F}}$, where $E_{\textrm{F}}$ is the Fermi energy. For dipole transitions, changes of $\Delta l=\pm1$ and $\Delta m_{l}=\pm1$ must occur, where $l$ and $m_{l}$ are the orbital and azimuthal orbital quantum numbers, respectively.  Since for $K$ absorption edges the orbital angular momentum is 0 and there is no spin-orbit splitting for the initial state, XMCD probes the magnetic polarization of the orbital moment of the final $p$ states. For $L$ edges, the spin and orbital polarization of the final $d$ states are probed.  Hence, XMCD can be viewed as an element- and orbital-specific probe of the magnetization of a sample \cite{Haskel_2005}.

Thin ($\sim\!100~\mu$m) single crystals of Ba$_{0.6}$K$_{0.4}$Mn$_{2}$As$_{2}$ were prepared using the self-flux solution-growth technique described elsewhere \cite{Lamsal_2013} and the chemical composition was determined by wavelength-dispersive x-ray spectroscopy \cite{Pandey_2013}.  Some single crystals were crushed for measurements on polycrystalline (powder) samples.  X-ray absorption spectra (XAS) and XMCD measurements were performed at the Advanced Photon Source at Argonne National Laboratory.  Measurements were made across the As and Mn $K$ x-ray absorption edges using station 4-ID-D, and measurements across the Mn $L_{2}$ and $L_{3}$ absorption edges, using a soft x-ray beam, were accomplished at station 4-ID-C.  The single-crystal and powder samples were mounted on Kapton or carbon tape and placed in a He flow cryostat to be cooled down to either $T=1.8$~K (4-ID-D) or 50~K (4-ID-C).  Magnetic fields of $H=3$~kOe were applied using either a superconducting magnet or an 8-pole electromagnet.  This low field value was chosen to saturate the FM signal while minimizing field-induced canting of the Mn moments (see Ref. \onlinecite{Pandey_2013}).  At both stations, data for circularly-polarized x-rays with helicity parallel [$\mu^{+}(E)$] and antiparallel to the beam direction [$\mu^{-}(E)$] were measured to obtain the XMCD signal which is  $(\mu^{+}-\mu^{-})$.  The XAS signal is proportional to $\frac{1}{2}(\mu^{+}+\mu^{-})$.  To eliminate experimental artifacts, data were taken for both the magnetic field applied parallel and antiparallel to the beam direction.  At 4-ID-D, a diamond phase retarder was used to circularly polarize the incoming beam, and the x-ray transmission through the samples was measured using an ion chamber detector.  Measurements were made using a lock-in technique which alternated the beam's helicity at a frequency of 13.1~Hz \cite{Suzuki_1998}.  At station 4-ID-C, the circular polarization of the beam was controlled by the undulator, and the total electron yield (TEY) and total fluorescence yield (TFY) were both measured \cite{Lang_2003}.

\begin{figure}[]
\centering
\includegraphics[width=1.0\linewidth]{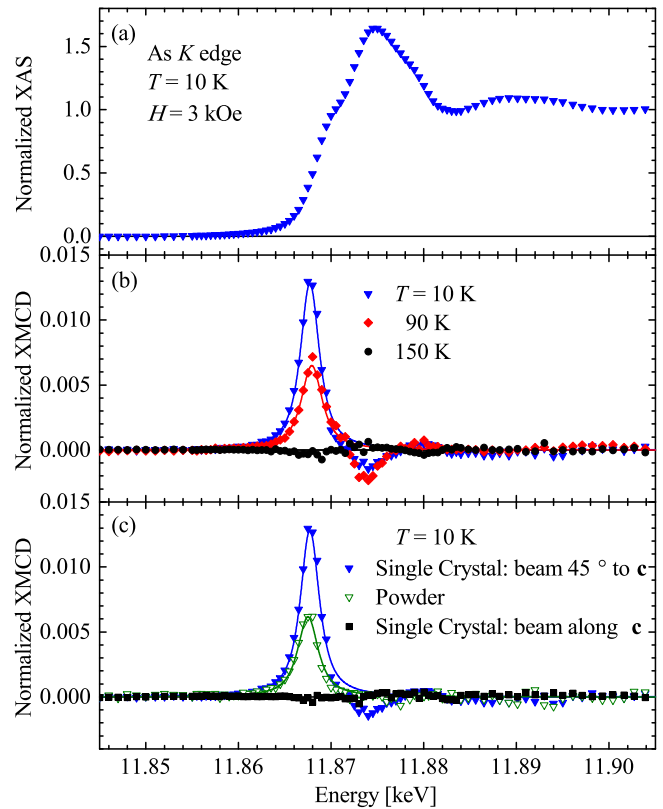}
\caption{(Color online) (a) The normalized XAS signal across the As $K$ edge from transmission measurements made on a single-crystal sample.  The XAS signal has been normalized to 1 at energies well above the edge.  (b)~Edge-step normalized XMCD data for $H=3$~kOe and various temperatures taken with the crystal's \textbf{c} axis oriented $45\degree$~away from the incoming beam ($H_{\textrm{\textbf{ab}}}=2.1$~kOe).  The $T=10$~K data relate to the XAS data in (a).  (c) Edge-step normalized XMCD data for $T=10$~K and $H=3$~kOe taken across the As $K$ edge for powder and single-crystal samples.  Data for the single crystal are shown for the \textbf{c} axis oriented either at $45\degree$~to the incoming beam or parallel to the beam.  The colored lines in (b) and (c) are fits to Lorenztian lineshapes.   \label{Fig1}}
\end{figure}

\begin{figure}[]
\centering
\includegraphics[width=1.0\linewidth]{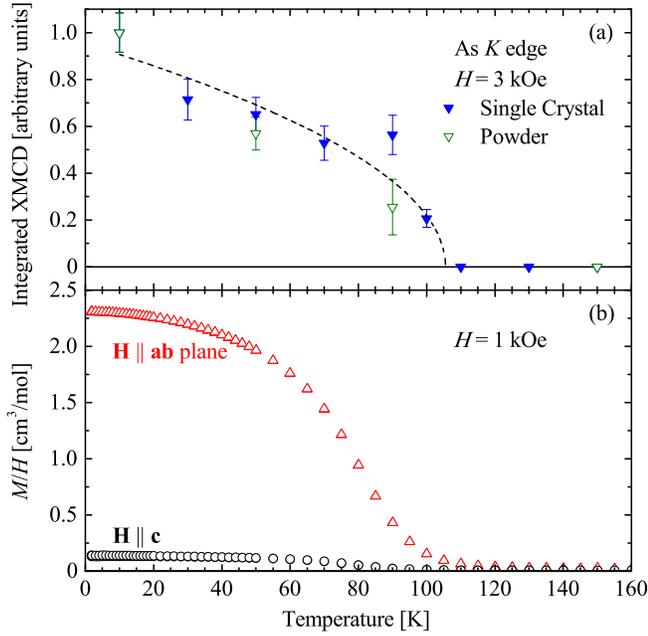}
\caption{(Color online)  (a) The temperature dependence of the integrated XMCD signal for the $H=3$~kOe As $K$ edge XMCD data.  Data are shown for both a single-crystal sample with its \textbf{c} axis oriented $45\degree$~away from the incoming beam and a powder sample.  The integrated XMCD signal has been determined from fits to the peak in the XMCD data with a Lorentzian lineshape.  Both datasets are normalized to their values for $T=10$~K, and the dashed line is a guide to the eye. (b) $\chi=M/H$ data for Ba$_{0.6}$K$_{0.4}$Mn$_{2}$As$_{2}$ from Ref. \onlinecite{Pandey_2013}.  The nonzero value of $\chi$ as $T\rightarrow0$~for $\textrm{\textbf{H}}$  applied parallel to the \textbf{c} axis is due to a slight  misalignment of the crystal ($\approx4\degree$) with respect to the field direction \cite{Pandey_2013}.  \label{Fig2}}
\end{figure}

Figure~\ref{Fig1}(a) shows the normalized XAS signal across the As $K$ edge for $H=3$~kOe and Fig.~\ref{Fig1}(b) shows the corresponding  XMCD signal normalized by the absorption edge step.  These data were obtained from transmission measurements made on a single crystal with its \textbf{c} axis directed $45\degree$~away from the incoming beam, so that $\textrm{\textbf{H}}$ has components along both the \textbf{c} axis and basal plane of the tetragonal lattice ($H_{\textrm{\textbf{c}}}=H_{\textrm{\textbf{ab}}}=2.1$~kOe).  For $T=10$~K and 90~K, a positive peak is observed just at the absorption threshold.  For $T=90$~K, the peak has decreased in magnitude and, for $T=150$~K, the peak is absent. The lines correspond to fits to the peak using a Lorentzian lineshape. These data provide evidence of the FM orbital polarization of the As $4p$  conduction band and are similar in appearance to previous As $K$ edge XMCD spectra in, for example, the Ga$_{1-x}$Mn$_x$As FM semiconductor \cite{Freeman_2008, Wadley_2010}.

The small negative peak observed at a slightly higher incident beam energy arises from linear dichroism contributions due to the incomplete circular polarization of the beam \cite{Lovesey_1996}.  This is supported by the absence of this negative peak in the data for the powder sample, which are plotted as green open triangles in Fig.~\ref{Fig1}(c).  Figure~\ref{Fig1}(c) also shows that for $T=10$~K (below $T_{\textrm{C}}$)  the positive peak in the XMCD signal is absent with the \textbf{c} axis  of the sample oriented parallel to the incident x-ray beam (black squares), whereas it is clearly evident for the $45\degree$ incident angle (blue triangles) as well as in the powder data.  Since XMCD is sensitive to the projection of the sample's magnetization along the beam direction, this demonstrates that the FM moments are perpendicular to \textbf{c} and lie within the basal plane.

\begin{figure}[]
\centering
\includegraphics[width=0.9\linewidth]{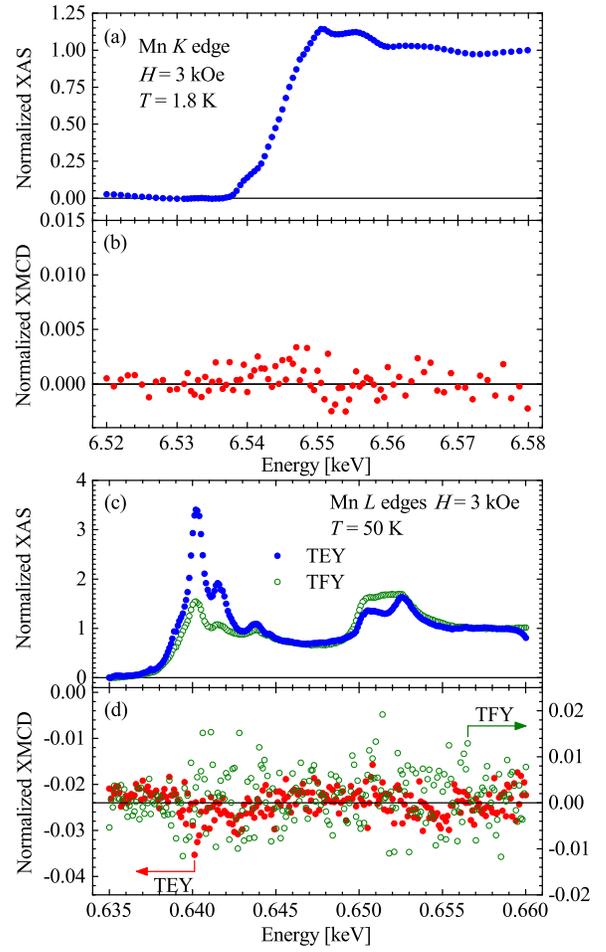}
\caption{(Color online)  (a)  The normalized XAS signal across the Mn $K$ edge for $T=1.8$~K and $H=3$~kOe from transmission measurements made on a single-crystal sample with its \textbf{c} axis oriented $45\degree$ away from the incoming beam.  (b) Normalized XMCD data corresponding to the XAS data in (a). (c) Normalized XAS data for $T=50$~K and $H=3$~kOe taken across the Mn $L_{2}$ and $L_{3}$ edges from measurements of the total electron yield (TEY) and total fluorescence yield (TFY) made on a single-crystal sample with its \textbf{c} axis  oriented $70\degree$ away from the incoming beam  ($H_{\textrm{\textbf{ab}}}=2.8$~kOe).  (d) Normalized XMCD data corresponding to the TEY (left axis) and TFY (right axis) data in (c).  The small constant offset in the TEY data is due to leakage currents and currents induced by external electromagnetic fields in the leads connected to the picoammeter used for the measurements.  The TFY data in (c) and (d) have not been corrected for absorption of the fluoresced x-rays by the sample \cite{footnote_1}. \label{Fig3}}
\end{figure}

The As $K$ edge XMCD spectra were recorded at several temperatures between $T=10$~and 150~K for both the powder and single-crystal samples. The areas determined from the fits to the Lorentzian lineshapes are taken as the integrated XMCD signal and are plotted as a function of temperature in Fig.~\ref{Fig2}(a). Below $T_{\textrm{C}}$, the data monotonically increase with decreasing temperature down to the base temperature for our As $K$ edge measurements.   Figure~\ref{Fig2}(b) shows the bulk magnetization data taken from Ref.~\onlinecite{Pandey_2013} and demonstrates that the temperature dependence of the integrated XMCD signal for both the single-crystal and powder samples agrees well with the temperature dependence of the bulk magnetization data for $\textrm{\textbf{H}}$ parallel to the \textbf{ab} plane.  The presence of the linear dichroism contamination in the XMCD signal for the single-crystal sample prevents further quantitative analysis of these data.

It is important to consider whether the measured FM can be associated with the Mn moments.  To this end, Fig.~\ref{Fig3} displays data from XAS and XMCD measurements on single-crystal samples for the Mn $K$ [Figs.~\ref{Fig3}(a) and \ref{Fig3}(b)] and $L_{2}$ and $L_{3}$ edges [Figs.~\ref{Fig3}(c) and \ref{Fig3}(d)].  For the Mn $L_{2}$ and $L_{3}$ edge measurements, TEY and TFY data were taken at $T=50$~K and the sample's  \textbf{c} axis  was oriented $70\degree$~away from the direction of the incident beam  ($H_{\textrm{\textbf{ab}}}=2.8$~kOe).  Measurements at the Mn $L_{2}$ and $L_{3}$ edges were also made on a powder sample and yielded similar data (not shown).  Hence, within the uncertainty of the measurement, we observe no features in the XMCD signals at the Mn $K$, $L_{2}$ and $L_{3}$ edges.

Using the normalized XMCD data at the Mn $L_{2}$ and $L_{3}$ edges from measurements performed on the FM semiconductor Ga$_{0.93}$Mn$_{0.07}$As at the same beam-line \cite{Keavney_2003}, we can estimate the upper limit for the possible Mn moment within the \textbf{ab} plane.  From Ref.~\onlinecite{Keavney_2003}, a FM ordered Mn moment of $3~\mu_{\textrm{B}}$ gave rise to a normalized XMCD signal of 2.  Applying this proportion to the spread of the noise in our TEY data (which is $\sim0.01$), we estimate that our experiments could detect a Mn moment $\geq0.02~\mu_{\textrm{B}}$.  Performing the same analysis using results from experiments made on La$_{0.7}$Sr$_{0.3}$MnO$_{3}$ ultra-thin films at the European Synchrotron Radiation Facility leads to the same conclusion \cite{Aruta_2009}.  Hence, within our experimental sensitivity of $0.02~\mu_{\textrm{B}}$, we detect no FM contribution from the Mn.  This result is also consistent with an upper limit of $M_{\textrm{\textbf{ab}}}=0.93(1)\times10^{-3}~\mu_{\textrm{B}}/\textrm{Mn}$ due to canting of the AFM Mn moments that we estimate from magnetization data reported in Ref.~\onlinecite{Pandey_2013}, as described in the Supplemental Material \cite{SM}.  We note that our TFY data may be strongly distorted by self-absorption effects \cite{footnote_1}, and we therefore use our TEY data as a basis for the upper estimate of  $0.02~\mu_{\textrm{B}}$ for the size of any FM ordered Mn moment present.  For measurements at the Mn $K$ edge, comparison to similar data taken for (In,Ga,Mn)As \cite{Freeman_2008} show that our experiments would detect a Mn moment $\geq0.15~\mu_{\textrm{B}}$.  This indicates that measurements at the Mn $L_{2}$ and $L_{3}$ edges are an order of magnitude more sensitive to the size of the moment than measurements at the Mn $K$ edge.

Using the normalized XMCD signal obtained for the As $K$ edge, we can determine the value of the orbital moment in the As $4p$ band by applying the sum rule  \cite{Lang_2003,Thole_1992,Guo_1996}
\begin{equation}
 -\frac{\langle L_{\textrm{z}} \rangle_{p}}{n_{p}^{\textrm{h}}}=\frac{\int_{E_{\textrm{F}}}^{E_{\textrm{c}}}(\mu^{+}-\mu^{-})dE}{P_{\textrm{c}}\int_{E_{\textrm{F}}}^{E_{\textrm{c}}} (\mu^{+}+\mu^{-}+\mu^{0})dE}.
\label{Eq1}
\end{equation}
Here, $\langle L_{\textrm{z}} \rangle_{p}$ is the $p$-projected orbital magnetic moment,  $n_{p}^{\textrm{h}}$ is the number of holes in the $p$ band, $(\mu^{+}-\mu^{-})$ is equal to the normalized XMCD data, $(\mu^{+}+\mu^{-}+\mu^{0})$ is $\frac{3}{2}$ times the normalized XAS data, $P_{\textrm{c}}$ is the degree of circular polarization of the beam, and $E_{\textrm{c}}$ is a cutoff energy above the edge.  The contribution to the normalized XAS data from unbound states near $E_{\textrm{F}}$ was modeled by a broadened step function \cite{Lang_2003} and subtracted before performing the integration in the denominator.  Using the $T=10$~K data for the powder sample, $P_{\textrm{c}}=0.95$, which is the nominal degree of circular polarization, and deconvolving the orientational average of the moments in the powder relative to the x-ray beam direction, we find that \mbox{$\frac{\langle L_{\textrm{z}} \rangle_{p}}{n_{p}^{\textrm{h}}}=-0.003(1)~\mu_{\textrm{B}}$}.  Taking $n_{p}^{\textrm{h}}=0.2~\textrm{holes}/\textrm{As}$, as done previously \cite{Pandey_2013, Glasbrenner_2013}, we find an As $4p$ orbital moment of $\langle L_{\textrm{z}} \rangle_{p}=-0.0006(2)~\mu_{\textrm{B}}/\textrm{As}$.  The negative sign for $\langle L_{\textrm{z}} \rangle_{p}$ implies that the As 4$p$ orbital moment is antiparallel to the total moment, which is presumably dominated by the As 4$p$ spin moment.
  
To investigate whether the size of the As 4$p$ orbital moment and its direction relative to the spin moment is reasonable in relation to the FM moment of $0.45(1)~\mu_{\textrm{B}}/\textrm{formula unit}$ determined from magnetization measurements \cite{Pandey_2013}, band structure calculations were performed with the Ba atom at the center of the body-centered tetragonal unit cell replaced by a K atom, corresponding to a K doping level of 50\%.  An artificial magnetic field was added to introduce a spin moment on the As sites, and local spin density approximation calculations were iterated to self-consistency.  The converged spin moments on the 2 As sites in the above supercell of the previous primitive cell are 0.24~$\mu_{\textrm{B}}/\textrm{As}$ and 0.21~$\mu_{\textrm{B}}/\textrm{As}$, and the orbital moments, induced via the spin-orbit interaction in the As 4$p$ band, are both \mbox{-0.001~$\mu_{\textrm{B}}/\textrm{As}$}. The magnitude of the orbital moments as well as their antiparallel orientations to the larger spin moments are in good agreement with the experimental results.

We also note that the value we obtain for the As $4p$ orbital moment is similar to the $4p$ orbital moments calculated for Fe and Ni \cite{Igarashi_1994}, and to those found from measurements at the As $K$ edge for thin films of the (Ga,Mn)As and (In,Ga,Mn)As FM semiconductors \cite{Freeman_2008, Wadley_2010}.   In these cases, an XMCD signal was also found at the $L_{2}$ and $L_{3}$ edges of the transition metals, and it was determined that the measured $K$ edge signals were due to polarization of the $4p$ orbital moments by the $3d$ moments of neighboring transition metal sites \cite{Freeman_2008, Wadley_2010, Igarashi_1994}.   As we have shown, this is not the case for Ba$_{0.6}$K$_{0.4}$Mn$_{2}$As$_{2}$ since no distinct peaks in the XMCD signal were observed at the Mn edges. Instead, the FM orbital polarization of the As $4p$ states originates from the spin-orbit interaction with spin-polarized As $4p$ states.  This assignment is consistent with our previous hypothesis that itinerant FM coexists with the Mn local moment AFM \cite{Pandey_2013}.

Summarizing, the presence of an XMCD signal at the As $K$ edge demonstrates that the FM order previously observed in Ba$_{0.6}$K$_{0.4}$Mn$_{2}$As$_{2}$ occurs in the As $4p$ conduction band.  Our results further show that the FM moments lie in the \textbf{ab} plane, perpendicular to the AFM ordered Mn moments.  Finally, the absence of a net XMCD signal (within the limits of our detection) at the Mn $K$, $L_{2}$, and $L_{3}$ edges shows that the FM is not associated with the Mn 3$d$ and 4$p$ bands, even in light of the substantial Mn-As hybridization evident from electronic structure calculations \cite{An_2009}.  These results demonstrate that the local-moment AFM order of the Mn and the itinerant FM in the As $4p$ conduction band are distinct, and illustrate the power of XMCD measurements to assign FM to a specific element and associated band.

\begin{acknowledgments}
We are grateful for discussions with P. Zajdel.  Work at the Ames Laboratory was supported by the Department of Energy, Basic Energy Sciences, Division of Materials Sciences \& Engineering, under Contract No. DE-AC02-07CH11358.  This research used resources of the Advanced Photon Source, a U.S. Department of Energy (DOE) Office of Science User Facility operated for the DOE Office of Science by Argonne National Laboratory under Contract No. DE-AC02-06CH11357.
\end{acknowledgments}

\newpage
\setcounter{equation}{0}
\setcounter{figure}{0}
\setcounter{table}{0}
\makeatletter
\renewcommand{\theequation}{S\arabic{equation}}
\renewcommand{\thefigure}{S\arabic{figure}}
\renewcommand{\bibnumfmt}[1]{[S#1]}
\renewcommand{\citenumfont}[1]{S#1}
\onecolumngrid
\begin{center}
\textbf{{\large Supplemental Materials:\\Itinerant ferromagnetism in the As 4$p$ conduction band of Ba$_{0.6}$K$_{0.4}$Mn$_{2}$As$_{2}$ identified by x-ray magnetic circular dichroism}}
\end{center}
\vspace{2ex}
\twocolumngrid

Since local-moment G-type antiferromagnetic (AFM) order of the Mn sublattice exists below $T_{\textrm{N}}=480$~K with an  ordered moment of $3.88(4)~\mu_{\textrm{B}}/\textrm{Mn}$ that lies along the \textbf{c} axis of the tetragonal unit cell \cite{Singh_2009,An_2009,Singh_2009b,Johnston_2011,Lamsal_2013}, it is important to consider how the magnetic field applied in the \textbf{ab} plane for the XMCD measurements would cant the ordered Mn moments away from the \textbf{c} axis.  Below, we set an upper limit of $0.93(1)\times10^{-3}~\mu_{\textrm{B}}/\textrm{Mn}$ for the magnetization in the \textbf{ab} plane that would occur due to canting of the AFM ordered Mn moments by a field of $H=2.8$~kOe.  This value is well below the estimated detection limit of $0.02~\mu_{\textrm{B}}$ for our XMCD measurements at the Mn  $L_{2}$ and $L_{3}$ edges.
\vspace{-0.05in}
\section{Expected Canting of the M\lowercase{n} Moments by the Applied Magnetic Field}

Previous results for insulating BaMn$_{2}$As$_{2}$ have shown that fits to the high-field range of magnetization versus field data accurately determine the magnetic susceptibility of the AFM ordered Mn sublattice \cite{Singh_2009}.  Hence, we expect that similar fits to data for metallic Ba$_{0.6}$K$_{0.4}$Mn$_2$As$_2$ may be used to estimate the susceptibility of the ordered Mn sublattice, provided that we consider data for either $T \ll T_{\textrm{C}}$ or $T_{\textrm{C}} \ll T \ll T_{\textrm{N}}$.

\begin{figure}[]
\centering
\includegraphics[width=1.0\linewidth]{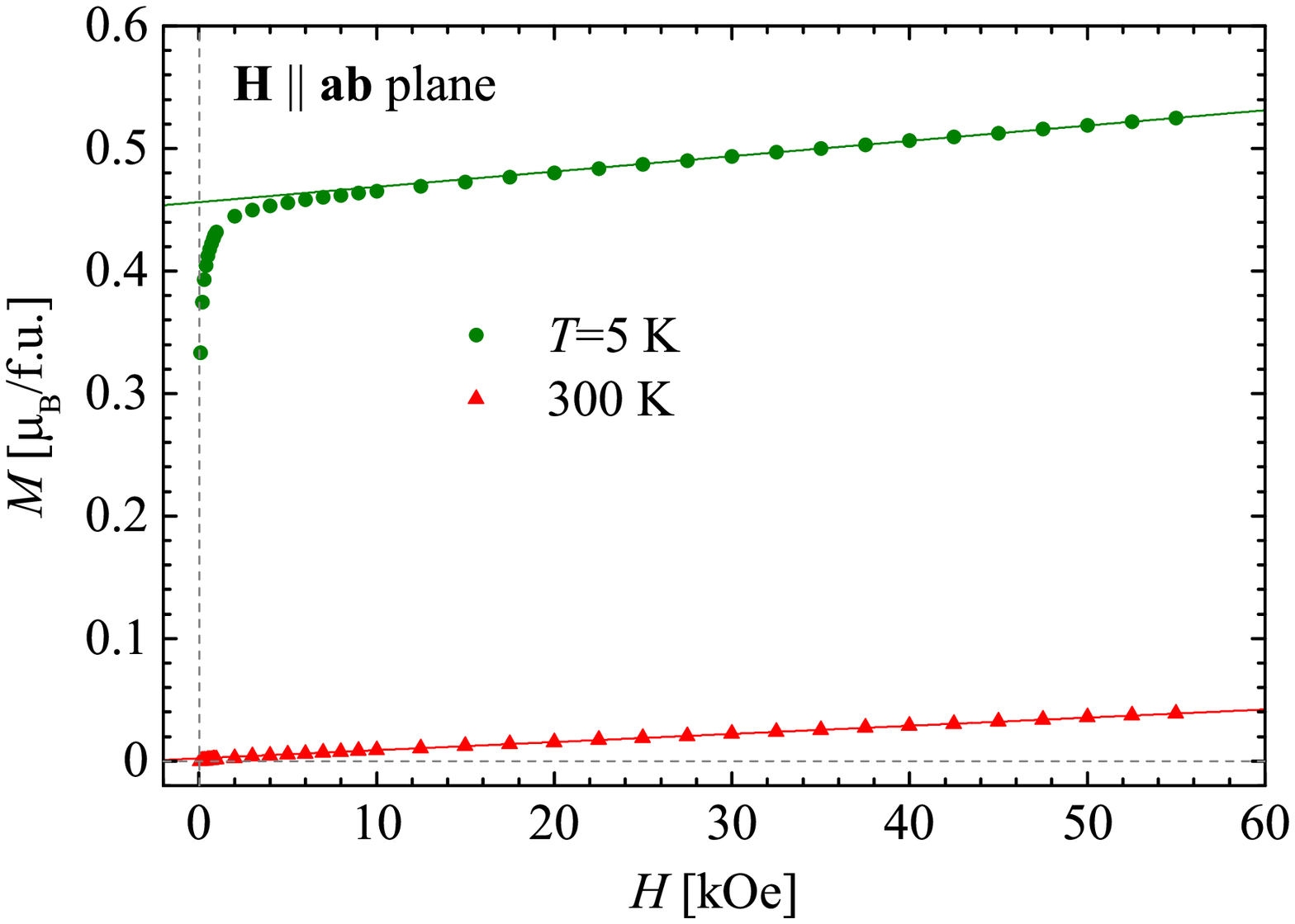}
\caption{(Color online)  Magnetization versus field data for a single crystal sample at various temperatures, taken from Ref.\ \onlinecite{Pandey_2013} .  The field was applied along the \textbf{ab} plane.  f.u. stands for formula unit.  Lines are linear fits between $H=30$ and 55~kOe.  \label{FigS1}}
\end{figure}

Figure \ref{FigS1} shows magnetization $M$ versus field data from measurements made on a single-crystal sample with \textbf{H} applied along the \textbf{ab} plane after cooling in zero field.  The $T=5$~K data illustrate the dramatic rise in $M$ at low fields due to the itinerant FM order.  Above $H\approx2$~kOe the contribution due to the itinerant FM order approaches saturation, and for  $H\geq30$~kOe $M$ is a linear function of $H$ up to at least  55~kOe.

By fitting the $T=5$~K data between $H=30$ and 55~kOe to a line, we find a slope of $\chi_{\textrm{\textbf{ab}}}=\frac{dM_{\textrm{\textbf{ab}}}}{dH}=1.25(1)\times10^{-3}~\mu_{\textrm{B}}/$kOe-f.u.  Multiplying this value by 2.8~kOe (the magnitude of the applied field in the \textbf{ab} plane during the Mn $L_{2}$ and $L_{3}$ edge XMCD experiments), and by assuming that the field-induced magnetization for $H\ge30$~kOe comes entirely from the Mn moments, we obtain an upper limit of $M_{\textrm{\textbf{ab}}}=1.75(1)\times10^{-3}~\mu_{\textrm{B}}/\textrm{Mn}$.  Applying the same procedure to the 300~K data (i.e.\ data for $T_{\textrm{C}}<<T<< T_{\textrm{N}}$), we obtain a slope of $\chi_{\textrm{\textbf{ab}}}=0.66(1)\times10^{-3}~\mu_{\textrm{B}}/$kOe-f.u. and a magnetization of $M_{\textrm{\textbf{ab}}}=0.93(1)\times10^{-3}~\mu_{\textrm{B}}/\textrm{Mn}$. Since the doped-hole FM may not be completed saturated at $H=3$~kOe, and $\chi_{\textrm{\textbf{ab}}}$ is likely somewhat temperature dependent, the discrepancy between the values of $M_{\textrm{\textbf{ab}}}$ for $T=5$ and 300~K appears reasonable.   In any case, both of the values are far below the detection limit of $\approx0.02~\mu_{\textrm{B}}$ for our XMCD experiments at the Mn $L_{2}$ and $L_{3}$ edges.


\begin{thebibliography}{99}
\bibitem{footnote_1} Attempts to correct the TFY data for self-absorption effects using two different methods were unsuccessful.
\end{thebibliography}
\end{document}